\documentclass[12pt,notitlepage]{article}
\usepackage{amsmath,amssymb}
\usepackage{graphicx}

\usepackage{color}

\def\ignore#1{{}}

\newcounter{sxn}

\newcounter{axn}

\newdimen\mybaselineskip
\newcommand{\beeq}{\begin{equation}}
\newcommand{\eneq}{\end{equation}}
\newcommand{\beqn}{\begin{eqnarray}}
\newcommand{\eeqn}{\end{eqnarray}}

\newcommand{\ba}{\begin{array}}
\newcommand{\ea}{\end{array}}

\newcommand{\be}{\begin{equation}}
\newcommand{\ee}{\end{equation}}
\newcommand{\bea}{\begin{eqnarray}}
\newcommand{\eea}{\end{eqnarray}}
\newcommand{\beal}{\setcounter{letter}{1} \begin{eqnarray}}
\newcommand{\eeal}{\addtocounter{equation}{1} \end{eqnarray}}

\newcommand{\larrow}{\,\,\,\,\hbox to 30pt{\rightarrowfill}
\,\,\,\,}
\newcommand{\slarrow}{\,\,\,\hbox to 20pt{\rightarrowfill}
\,\,\,}

\def\la{\raise.16ex\hbox{$\langle$}\lower.16ex\hbox{}  }
\def\ra{\, \raise.16ex\hbox{$\rangle$}\lower.16ex\hbox{} }

\def\psibar{ \psi \kern-.65em\raise.6em\hbox{$-$} \lower.6em\hbox{} }
\def\psibarb{ \psi \kern-.65em\raise.6em\hbox{$-$}  }

\def\Ai {\hbox{Ai}}
\def\Bi {\hbox{Bi}}

\begin{document}

\thispagestyle{empty}





\begin{center}  
{\LARGE \bf   On the Significance of Black Hole Quasinormal Modes: A Closer Look}

\vspace{1cm}

{\bf  Ramin G.~Daghigh$^{1,2}$, Michael D.\ Green$^3$, Jodin C.~Morey$^4$}
\end{center}

\centerline{\small \it $^1$ William I. Fine Theoretical Physics Institute, University of Minnesota,
	Minneapolis, Minnesota, USA 55455}
\vskip 0 cm
\centerline{} 

\centerline{\small \it $^2$ Natural Sciences Department, Metropolitan State University, Saint Paul, Minnesota, USA 55106}
\vskip 0 cm
\centerline{} 

\centerline{\small \it $^3$ Mathematics and Statistics Department, Metropolitan State University, Saint Paul, Minnesota, USA 55106}
\vskip 0 cm
\centerline{} 

\centerline{\small \it $^4$ School of Mathematics, University of Minnesota, Minneapolis, Minnesota, USA 55455}
\vskip 0 cm
\centerline{} 

\vspace{1cm}
\begin{abstract}
It is known that approximating the Regge-Wheeler potential with step functions significantly modifies the Schwarzschild black hole quasinormal mode spectrum. Surprisingly, this change in the spectrum has little impact on the ringdown waveform. We examine whether this issue is caused by the jump discontinuities and/or the piecewise constant nature of step functions.  We show that replacing the step functions with a continuous piecewise linear function does not qualitatively change the results.  However, in contrast to previously published results, we discover that the ringdown waveform can be approximated to arbitrary precision using either step functions or a piecewise linear function.  Thus, this approximation process provides a new mathematical tool to calculate the ringdown waveform. In addition, similar to normal modes, the quasinormal modes of the approximate potentials seem to form a complete set that describes the entire time evolution of the ringdown waveform.  We also examine smoother approximations to the Regge-Wheeler potential, where the quasinormal modes can be computed exactly, to better understand how different portions of the potential impact various regions of the quasinormal mode spectrum.

\end{abstract}

\newpage

\section{Introduction}

Quasinormal modes (QNMs) of black holes are the natural vibrational modes of perturbations in the spacetime exterior to a black hole.  QNM frequencies are discrete and complex.  The imaginary part of the frequency indicates the presence of damping, a necessary consequence of boundary conditions that require energy to be carried away from the system.  

QNMs play an important role in gravitational wave astronomy\cite{LIGO} because they determine the shape of the ringdown phase in a binary black hole merger and, consequently, provide clues to the nature of the postmerger object.  There have also been attempts to link the high overtone QNMs of black holes to the quantum structure of spacetime\cite{HighOvertone1, HighOvertone2, HighOvertone3}.  

The main goal of this paper is to better understand the connection between the black hole QNM frequency spectrum and the ringdown waveform by studying some alternatives to the Regge-Wheeler potential.

The axial (odd-parity) perturbations in a Schwarzschild spacetime in the linear approximation of general relativity are described by the Regge-Wheeler equation 
\beeq
\partial_t^2\psi_l+\left(-\partial_x^2+V_l\right)\psi_l=0~,
\label{RWE}
\eneq  
where $t$ is time, $l$ is the orbital angular momentum number and $x$ is the tortoise coordinate.  We use the geometric unit system where $G=c=1$.  The tortoise coordinate is linked to the radial coordinate, $r$, according to
\beeq
dx=\frac{dr}{1-\frac{r_{Sch}}{r}}~,
\label{tortoise}
\eneq
where $r_{Sch}$ is the Schwarzschild radius.  The integrated form is
\beeq
x=r+r_{Sch} \ln(r-r_{Sch})+\mbox{constant}~,
\label{tortoise1}
\eneq
where we usually choose the constant so that the maximum of the potential is at $x=0$.  $V_l$ is the Regge-Wheeler potential
\beeq
V_l(r)=\left(1-\frac{r_{Sch}}{r}\right) \left[\frac{l(l+1)}{r^2}+(1-s^2)\frac{r_{Sch} }{r^3}\right]~,
\label{RWP}
\eneq
where $s$ is the spin of the perturbation with values $0$, $1$ and $2$ for scalar, electromagnetic and gravitational fields respectively.   

If we assume the perturbations depend on time as $\psi_l(x,t)=e^{-i \omega t} \phi_l(x)$, we can write the Regge-Wheeler equation as
\beeq
\partial_x^2\phi_l+\left(\omega^2-V_l\right)\phi_l=0~,
\label{RWEnoTim}
\eneq  
where $\omega$ is the complex QNM frequency to be determined.  For simplicity, in the remainder of this paper, we choose units such that $r_{Sch}=1$.

With the above time-dependence, the boundary conditions at the event horizon and infinity are, respectively,
\beeq
\begin{array}{ll}
	\phi_l(x) \rightarrow e^{-i\omega x}  & \mbox{as $x \rightarrow -\infty$ ($r \rightarrow 1$)}~,\\
	\phi_l(x) \rightarrow e^{i\omega x}  & \mbox{as $x \rightarrow \infty$ ($r \rightarrow \infty$)}~.
\end{array}       
\label{B.C.}
\eneq

In \cite{Nollert}, Nollert shows that  approximating the Regge-Wheeler potential with a series of step potentials modifies the Schwarzschild black hole QNM spectrum significantly. He found that QNMs of the modified potential line up along the real axis instead of the imaginary axis as they do for a Schwarzschild black hole.  Surprisingly, this significant change in the QNM spectrum has little impact on the ringdown waveform. 

A similar phenomenon appears in the context of exotic compact objects (ECOs).  For a recent review article on these objects, see \cite{PaniReview}.  ECOs do not possess an event horizon.  As a result, the boundary condition at the surface of the ECO is not a purely ingoing wave.  It has been shown in \cite{Pani} that the change in the boundary conditions drastically affect the QNM spectrum.  For example, the QNMs for a traversable wormhole line up along the real axis in contrast to the Schwarzschild QNMs.  Interestingly, this drastic change in the complex QNM frequency spectrum does not affect the waveform in the early stages of the ringdown.  The change in the boundary conditions can only be detected by the appearance of echoes at later times in the ringdown.

In this paper, we explore the question of whether the change in the QNM spectrum that Nollert\cite{Nollert} observed is caused by the jump discontinuities and/or the piecewise constant nature of step functions.  We can avoid these issues, and still compute the solutions exactly, by using a continuous piecewise linear potential. Our question is whether this restores the original Schwarzschild QNM spectrum.  However, we show that this is not the case and our results remain qualitatively consistent with the previously used step functions.  In addition, we discover that the ringdown waveform can be approximated to arbitrary precision using either step functions or a piecewise linear function.

We also provide two smoother approximations to the Regge-Wheeler potential.  These approximate potentials are chosen so that they have the same asymptotic behavior as the Regge-Wheeler potential at the event horizon and infinity, but simple enough that one can determine their QNM spectrum exactly.    This provides a tool to gain a qualitative understanding of how different regions of the QNM spectrum are linked to different regions of the potential. Other authors also have explored alternative potentials, where the QNMs can be determined exactly/analytically.  For a partial list of these potentials, see \cite{Visser}. 

We structure the paper as follows. In Sec.\ \ref{Sec:RWlinear}, we
approximate the Regge-Wheeler potential using a continuous piecewise linear potential and we calculate the QNM spectrum and the ringdown waveform of this potential.  In Sec.\ \ref{Sec:OtherPotentials}, we introduce two smoother alternatives to the Regge-Wheeler potential and analyze their QNMs and rindown waveform.  In Sec.\ \ref{Sec:conclusions}, we provide a summary of the results with concluding remarks.

\section{Piecewise Linear Potential}
\label{Sec:RWlinear}

Our first approximation to the Regge-Wheeler potential is a piecewise linear function:
\beeq
V(x) = \left\{ \begin{array}{ll}
	0& \mbox{$x < x_0$}~\\ 
	\vdots ~\\  
    V_{i-1}+\dfrac{V_i-V_{i-1}}{x_i-x_{i-1}} (x-x_{i-1}) & \mbox{$x_{i-1} \le x < x_i$~} \\ 
    \vdots ~\\ 
	0	& \mbox{$x \ge x_N$}~,
\end{array}
\right.        
\label{Vlinear}
\eneq
for $i=1,2,3,\dots,N$ where $N$ is the number of line segments used. $V_i$ is the height of the Regge-Wheeler potential at $x_i$.  We choose $V_0=V_N=0$.

The solution to (\ref{RWEnoTim}) using potential (\ref{Vlinear}) is
\beeq
\phi(x) = \left\{ \begin{array}{ll}
    A e^{i\omega x} +B e^{-i\omega x} & \mbox{$x < x_0$}~\\  
	\vdots ~\\  
	C_i~ \Ai\left(\dfrac{-\omega^2+V_{i-1}+\dfrac{V_i-V_{i-1}}{x_i-x_{i-1}} (x-x_{i-1})}{\left(\dfrac{V_i-V_{i-1}}{x_i-x_{i-1}}\right)^{2/3}}  \right) \\
	~~~+	D_i~ \Bi\left(\dfrac{-\omega^2+V_{i-1}+\dfrac{V_i-V_{i-1}}{x_i-x_{i-1}} (x-x_{i-1})}{\left(\dfrac{V_i-V_{i-1}}{x_i-x_{i-1}}\right)^{2/3}}  \right) & \mbox{$x_{i-1} \le x < x_i$~} \\  
	\vdots ~\\  
    E e^{i\omega x} +F e^{-i\omega x}	& \mbox{$x \ge x_N$}~,
\end{array}
\right.        
\label{VlinSolution}
\eneq
where $\Ai(z)$ and $\Bi(z)$ are the two linearly independent Airy functions.  $A$, $B$, $C_i$, $D_i$, $E$ and $F$ are constants.  The boundary conditions (\ref{B.C.}) require that $A=F=0$.

The complex QNM frequencies are determined by imposing the following conditions 
\beeq
\begin{array}{ll}
	\phi_{L}(x)|_{x=x_j} = \phi_{R}(x)|_{x=x_j}~\\
	\phi_{L}'(x)|_{x=x_j} = \phi_{R}'(x)|_{x=x_j}~
\end{array}       
\label{DiscontinuityEq}
\eneq
at the points $x_j$ for $j=0,1,2,\dots,N$ where the potential $V(x)$ is non-differentiable, and $\phi_{L}$ and $\phi_{R}$ are the solutions immediately to the left and right of $x_j$.  Here, prime indicates differentiation with respect to $x$.

In Table I, we show the complex QNM frequencies of four piecewise linear functions fitted to the Regge-Wheeler potential for a scalar perturbation with $l=2$ ($V_{l=2}^{scalar}$).  These potentials have $N= 3, 5, 6, 7$ line segments supported on the domain $[-4, 8]$, $[-4, 16]$, $[-4, 20]$ and $[-8, 20]$ respectively. Each  line segment has width $\Delta x=x_i-x_{i-1}=4$.  The choice of the domain of the approximate potentials is made in a way to fit the Regge-Wheeler potential as well as possible with the specified number of line segments.   For comparison, we also include the first eleven complex QNM frequencies for $V^{scalar}_{l=2}$, which we calculated using Leaver's continued fraction method\cite{Leaver} with Nollert's improvement\cite{Nollert1}. The roots found using this method are consistent with those found by other techniques.  See, for example, the roots calculated in \cite{MatrixMethod} using both a Matrix and a sixth order WKB method.

\vspace{0.5cm}
\footnotesize
\begin{tabular}{cccccc}
	\multicolumn{6}{c}{Table I:  QNMs of piecewise linear potentials fitted to $V^{scalar}_{l=2}$ with $\Delta x=4 $} \\ 
	\hline
	$n$ & $N=3$ &  $N=5$ & $N=6$  &  $N=7$  & $V_{l=2}^{scalar}$  \\ 
	\hline 
	0 & $0.7510- 0.1873i$ &  $0.4084 - 0.1243i$ & $0.3230 - 0.0966i$ & $0.3230- 0.0966i$ & $0.967288-0.193518i$\\ 
	1 & $0.9563- 0.1457i$ &  $0.6172 - 0.1523i$ & $0.4992-0.1224i$ & $0.4992 - 0.1224i$ & $0.927701-0.591208i$   \\ 
	2 & $1.0806 - 0.2918i$ &  $0.7792 - 0.1461i$ & $0.6597 - 0.1308i$ &$0.5444- 0.2860i$ &$0.861088-1.017117i$ \\ 
	3 & $1.2799- 0.3469i$ & $0.9424 - 0.1359i$ & $0.7850 - 0.1318i$  &$0.6529- 0.1372i$ &$0.787726-1.476193i$ \\ 
	4 & $1.4995- 0.4011i$ &   $1.0191 - 0.1988i$  & $0.9246 - 0.1358i$ &$0.7851- 0.1319i$ &$0.722598-1.959843i$\\ 
	5 & $1.7095- 0.4297i$ &    $1.1292 - 0.2472i$  &  $0.9969 - 0.1572i$ &$0.9264- 0.1347i$ &$0.669799-2.456822i$\\ 
	6 & $$ &  $1.2629 - 0.2527i$ & $1.0999 - 0.2187i$  &$0.9786- 0.1434i$ &$0.627772-2.959909i$\\ 
	7 & $$ &   $1.4152 - 0.2718i$ &  $1.1885- 0.2218i$  &&$0.593941-3.465522i$\\ 
	8 & $$ &   $1.5665 - 0.2946i$ & $1.3197 - 0.2285i$ &&$0.566173-3.972018i$\\ 
	9 & $$ &   $1.7122 - 0.3175i$ & $1.3909 - 0.2547i$ &&$0.542926-4.478663i$\\ 
	10 & &   $1.8573 - 0.3297i$ & $1.4473 - 0.2426i$  &&$0.523115-4.985130i$\\ 
	\label{Table1}
\end{tabular} 
\normalsize

\begin{figure}[th!]
	\begin{center}
		\includegraphics[height=5.cm]{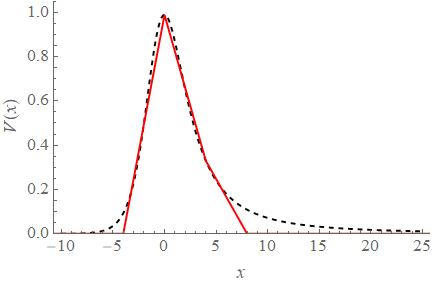}
		\includegraphics[height=5.cm]{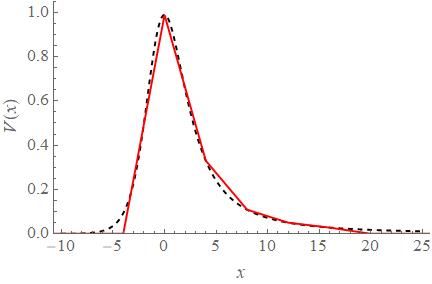}
	\end{center}
\vspace{-0.7cm}
	\caption{\footnotesize $N=3$ (left) and $N=6$ (right) piecewise linear potentials, shown in solid red, with line segments of width $\Delta x=4 $ supported on the domain $[-4, 8]$ and $[-4, 20]$ respectively.  For comparison, we plot $V_{l=2}^{scalar}$ in dashed black.}
	\label{Vlin-graph}
\end{figure}

In Figure \ref{Vlin-graph}, we plot the approximations using $N=3$ and $N=6$ along with the potential $V^{scalar}_{l=2}$ to show how well they match.   Note that the Regge-Wheeler potential can be approximated very well with only $N=6$ line segments.

\begin{figure}[th!]
	\begin{center}
		\includegraphics[height=10cm]{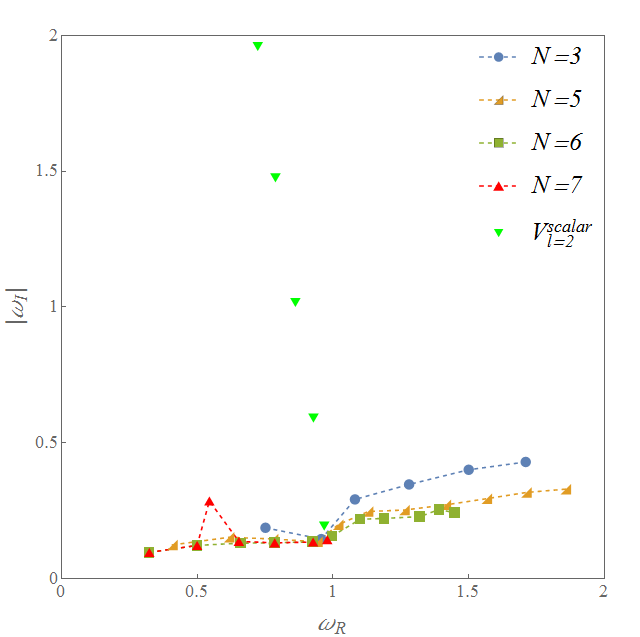}
	\end{center}
	\caption{\footnotesize    QNMs , provided in Table I, for  $N=3, 5, 6, 7$ piecewise linear potentials fitted to $V_{l=2}^{scalar}$.  For comparison, we also show QNMs of $V_{l=2}^{scalar}$.  }
	\label{Vlin-datagraph}
\end{figure}

In Figure \ref{Vlin-datagraph}, we plot the data in Table I.  As one can see, QNM frequencies of the potential (\ref{Vlinear}) line up along the real axis while the QNM frequencies of the Regge-Wheeler potential line up along the imaginary axis.  Note that we do not provide eleven roots for $N=3$ and $7$.  We are unable to find the higher overtone QNMs, with $|\omega| \gtrsim 2$, due to the increasingly oscillatory behavior of the Airy functions.  In addition, larger values of $N$ make the numerical calculations more challenging since this generates more Airy functions in the solution.  Other techniques need to be employed to find higher overtones.  The data in Table I show how the QNM spectrum changes with the addition of new line segments.  One might expect that as $N$ increases, the QNMs would get closer to those of the Regge-Wheeler potential.  Instead, the QNMs of the potential (\ref{Vlinear}) get closer to the real axis.  

For $N=7$, one of the data points has a significantly higher imaginary component.  Similar unusual data points also appear when the Regge-Wheeler potential is approximated with step functions as seen in FIG.\ 2 of \cite{Nollert}.  For larger values of $N$, Nollert found more of these unusual data points with even larger damping (i.e. larger $|\omega_I|$). We do not know if there is a discernible trend to these unusual data points as $N$ increases, but it may be worth further investigation. 

To generate the QNM ringdown waveform, we numerically solve the Regge-Wheeler wave equation (\ref{RWE}) using the initial data
\beeq
\psi^{scalar}_{l=2}(x,0)={\cal A} \exp \left(- \frac{(x-x_0)^2}{2\sigma^2} \right),~  \partial_t \psi^{scalar}_{l=2}|_{t=0}=-\partial_x \psi^{scalar}_{l=2}(x, 0)~,
\label{GaussianWave}
\eneq  
where we use $\sigma=1$, $x_0=-40$ and ${\cal A}=30$.  We choose the observer to be located at $x=90$.   To carry out the calculations, we use the built-in Mathematica commands for solving partial differential equations.

\begin{figure}[th!]
	\begin{center}
		\includegraphics[height=5cm]{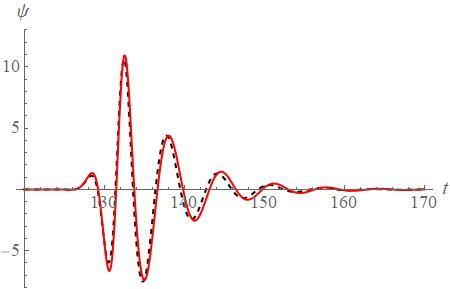}
		\includegraphics[height=5cm]{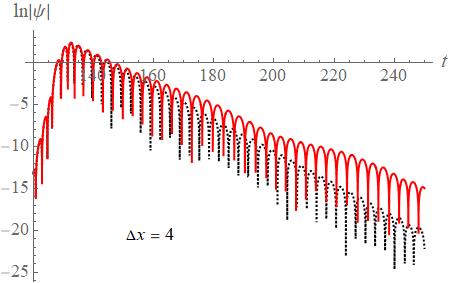}
		\includegraphics[height=5cm]{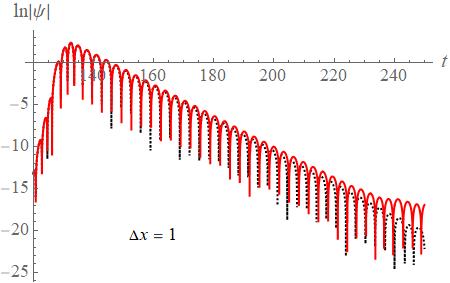}
		\includegraphics[height=5cm]{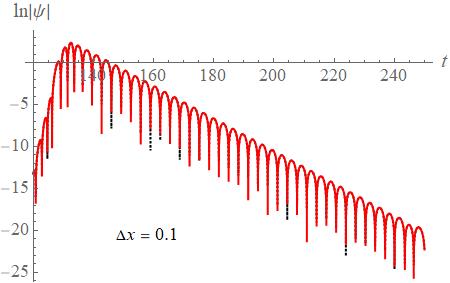}
	\end{center}
	\caption{\footnotesize The upper two graphs show, in solid red, $\psi$ and $\ln |\psi|$ as a function of time for $N=6$ piecewise linear potential with $\Delta x=4$ on the domain $[-4,20]$.  
	The lower two graphs show the ringdown waveform, in solid red, for piecewise linear potentials on $[-24, 250]$ with $\Delta x=1 $ (left) and $\Delta x=0.1$ (right).  In all graphs, for comparison, the ringdown waveform for $V_{l=2}^{scalar}$ is plotted in dashed black. 
    }
	\label{Vlin-ringdown}
\end{figure}

At the top of Figure \ref{Vlin-ringdown}, we provide the ringdown waveform $\psi$, as a function of time, for the potential (\ref{Vlinear}) with $N=6$.  For comparison, we also provide the ringdown waveform caused by the Regge-Wheeler potential  $V^{scalar}_{l=2}$ in dashed black.  The ringdown waveform is nearly the same for both potentials.  However, more details can be observed in the plot of $\ln|\psi|$ to the right, where we notice the ringdown is less damped for the piecewise linear potential. In addition, we observe some echo-like behavior (bumps) at a later time in the plot of the logarithm.

At this point, our results are consistent with Nollert's results\cite{Nollert} where he uses eight step functions to approximate the Regge-Wheeler potential for gravitational perturbations ($s=2$) with $l=2$.   However,  things change when we use a larger number of line segments.  In the lower graphs of Figure \ref{Vlin-ringdown}, we show the ringdown waveform for piecewise linear potentials, with $\Delta x=1$ and $0.1$ respectively, fitted to $V_{l=2}^{scalar}$ on the domain  $[-24,250]$.   In the lower left graph, we can see the two ringdown waveforms match almost perfectly initially and they diverge later.  The difference in the ringdown waveform disappears when we reduce the width to $0.1$.  This is a peculiar situation.  We have two potentials with completely different QNM spectra that produce identical ringdown waveforms in the limit where $\Delta x$ of the piecewise linear potential approaches zero.

\begin{figure}[th!]
	\begin{center}
		\includegraphics[height=5cm]{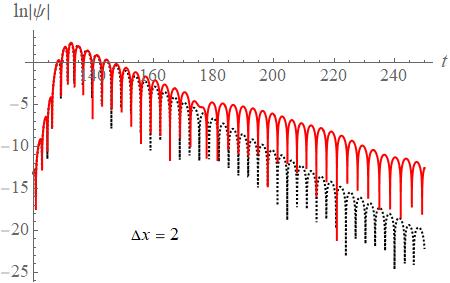}
		\includegraphics[height=5cm]{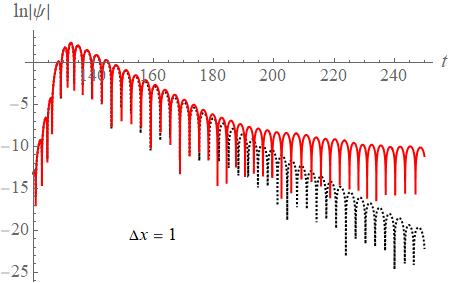}
		\includegraphics[height=5cm]{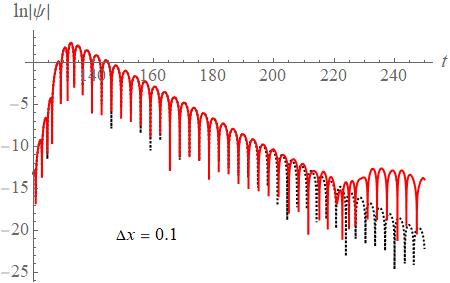}
		\includegraphics[height=5cm]{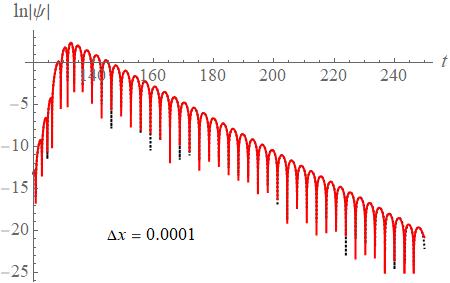}
	\end{center}
	\caption{\footnotesize The upper left graph shows the ringdown waveform, in solid red, for a  potential constructed from $12$ step functions with a width of $\Delta x=2 $ supported on the domain $[-4, 20]$. In the upper right graph, $\Delta x=1 $ on $[-24, 250]$.  In the lower left graph, $\Delta x=0.1$ on $[-24, 250]$. Finally, in the lower right graph, $\Delta x=0.0001$ on $[-24, 250]$. In all graphs, for comparison, the ringdown waveform for $V_{l=2}^{scalar}$ is plotted in dashed black.}
	\label{Vstep-ringdown}
\end{figure}

The same situation also happens when we approximate the Regge-Wheeler potential using step functions.  In Figure \ref{Vstep-ringdown}, we compare the ringdown waveform of four potentials with a various number of step functions to the Regge-Wheeler waveform.  In the upper left graph, our potential is constructed from $12$ step functions with a width of $\Delta x=2$ supported on the domain $[-4, 20]$.
In the upper right graph, we use $\Delta x=1 $ on $[-24, 250]$.  In the lower left graph, we use $\Delta x=0.1 $ on $[-24, 250]$. Finally, in the lower right graph, $\Delta x=0.0001 $ on $[-24, 250]$. 

For the case $\Delta x=2 $, we observe lower damping compared to the Regge-Wheeler waveform.  However, as $\Delta x$ becomes smaller, the waveform of the potential with step functions converges to the Regge-Wheeler waveform.   Note that to achieve convergence, we need a smaller $\Delta x$ for step versus linear functions.  This is not surprising, since with the same number of pieces a piecewise linear function fits the  Regge-Wheeler potential more accurately than step functions. 

Nollert\cite{Nollert} was able to determine the QNM frequency spectrum for potentials with up to $2084$ step functions.  In all cases, he found the complex QNM frequencies lined up along the real axis in contrast to the QNM frequencies for the Regge-Wheeler potential that line up along the imaginary axis.

\begin{figure}[th!]
	\begin{center}
		\includegraphics[height=4.9cm]{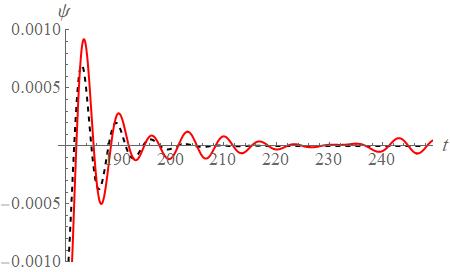}
		\includegraphics[height=4.9cm]{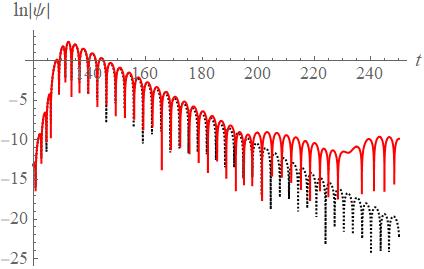}
		\includegraphics[height=8cm]{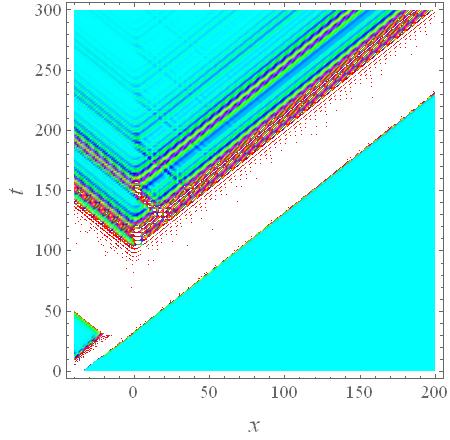}
	\end{center}
	\vspace{-.5cm}
	\caption{\footnotesize  The ringdown waveform for a piecewise linear potential on $[-24,  252]$, where we divide the domain into two intervals.   On $[-24,  32]$, where the potential changes rapidly, $\Delta x=1$ and on $[32,  252]$ $\Delta x=20 $.   The upper left graph shows $\psi$, for an observer at $x=90$, at very late times in solid red.   The upper right graph shows $\ln |\psi|$ as a function of time in solid red.  For comparison, the ringdown waveform for $V_{l=2}^{scalar}$ is shown in dashed black. The lower graph is a density plot of the evolution of the scattered wave, $\psi$, as a function of time and position $x$.  Here the echoes can be clearly seen as originating from the transition points between each line segment of the potential, due to the reflection of part of the incoming wave, and then reflecting back from the potential at $x=0$}
	\label{3Decho}
\end{figure}

For a small number of step functions, our results shown in Figure  \ref{Vstep-ringdown} (upper left graph) seem to agree with the results of \cite{Nollert} that the ringdown waveform of the approximate potential composed of  step functions is less damped than the Regge-Wheeler potential. According to Nollert's analysis, this is because the fundamental QNM for the approximate potential is less damped than the fundamental QNM of the Regge-Wheeler potential.  By increasing the number of step functions, we would expect to see less damping since the QNMs get closer to the real axis\cite{Nollert}.  However, what we observe in Figure \ref{Vstep-ringdown} is the opposite.  In fact, with a sufficient number of step functions, we can produce to arbitrary accuracy the same ringdown waveform as in the Regge-Wheeler case.  Of course, if the QNMs of the step potentials form a complete set, they should in principle be able to generate any waveform.

To better understand why the ringdown waveform appears to be less damped when we use large $\Delta x$, in Figure \ref{3Decho} we plot the ringdown waveform for a piecewise linear potential on $[-24,  252]$, where we divide our domain into two intervals.   On $[-24,  32]$, where the potential changes rapidly, we use $\Delta x=1$ and on $[32,  252]$ we use $\Delta x=20 $. In the upper two graphs of Figure \ref{3Decho}, we show the ringdown behavior of the constructed piecewise linear potential where the echoes are visible.  For comparison, we also show the ringdown waveform for $V_{l=2}^{scalar}$ in dashed black. Note that the ringdown waveform of the piecewise linear potential agrees well with the ringdown waveform of $V_{l=2}^{scalar}$ at early stages.  This is because we use a small $\Delta x$  on $[-24,  32]$ where the bulk of our potential is located.

In the lower graph of Figure \ref{3Decho}, perpendicular to the transmitted Gaussian wavepacket (which is moving to the right as $t$ increases) we see reflected waves (moving to the left) originating from the transition points between each line segment of the potential.  Observe that the distance between the reflected waves can be measured to be $20$, consistent with the width of the line segments.  The reflected waves then bounce back from the potential at $x=0$ and appear in the waveform at later times in the form of echoes.  The first echo originates from $x=52$ where the first transition between line segments with $\Delta x=20$ occurs.

For smaller $\Delta x$, these echoes cannot be easily distinguished from the waveform, but add to the amplitude, causing it to appear less damped.  When $\Delta x$ becomes very small, the change in slope between the line segments in the potential is small enough that the echoes are too small to contribute to the wave in any significant way.  The same argument can be made for potentials with step functions used in \cite{Nollert}.


\section{Smoother Approximate Potentials}
\label{Sec:OtherPotentials}

We consider two simple potentials, which have the same asymptotic behavior as the Regge-Wheeler potential as $r\rightarrow 1$ and $r\rightarrow \infty$.  These potentials are shown in Figures \ref{VIgraph} and \ref{VIIgraph}.  For comparison, on the same graphs, we show the Regge-Wheeler potential for scalar field perturbations with $l=2$.  Details are provided below.

\subsection{Potential I}

\begin{figure}[th!]
	\begin{center}
		\includegraphics[height=7.5cm]{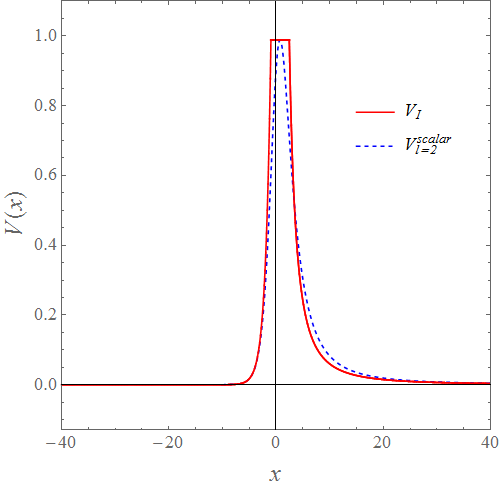}
		\includegraphics[height=8.5cm]{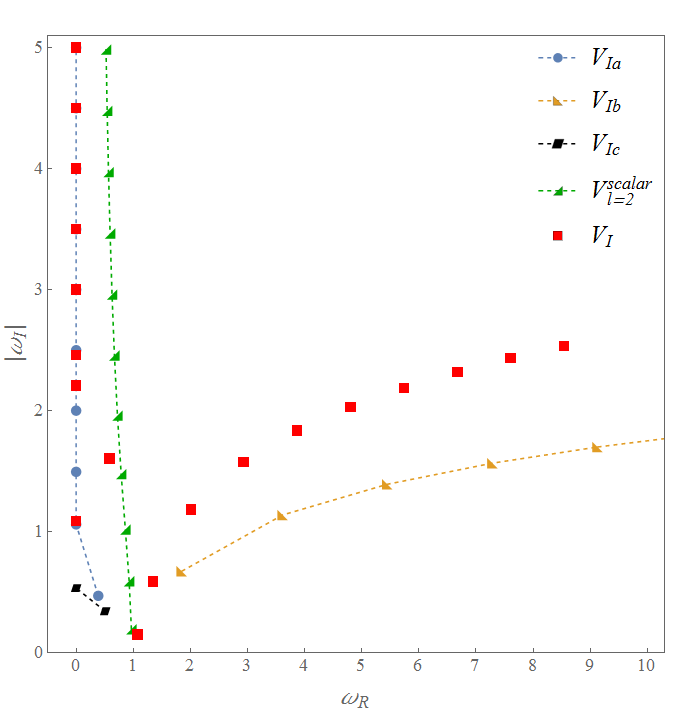}
	\end{center}
	\vspace{-0.5cm}
	\caption{\footnotesize Potential $V_I$ is plotted on the left with a solid red line.  The non-differentiable points are located at $x_0=-0.9575$ and $x_1=2.4637$. For comparison, Regge-Wheeler Potential for $V_{l=2}^{scalar}$ is plotted with a dashed blue line. The QNM data of Table II is presented on the right.  For comparison, the QNM spectrum of $V_{l=2}^{scalar}$ is also included. }
	\label{VIgraph}
\end{figure}

We construct the first potential with two functions that have the same asymptotic behavior as the Regge-Wheeler potential as $x\rightarrow \pm \infty$. We then connect these two functions with a straight horizontal line as shown in Figure \ref{VIgraph}.  The constructed potential is:
\beeq
V_{I}(x) = \left\{ \begin{array}{ll}
	 \left[l(l+1)+(1-s^2)\right]e^{x-1}  & \mbox{$x < x_0$}~\\  \\
	 V_{max}  & \mbox{$x_0 \le x < x_1$}~\\  \\
	 l(l+1)/x^2  & \mbox{$x \geq x_1$~,}
\end{array}
\right.        
\label{VIpotential}
\eneq
where $V_{max}$ is the height of the Regge-Wheeler potential. To have simpler equations, we choose the constant in Eq.\ (\ref{tortoise1}) to be zero for this potential.  For that reason, $V_{max}$ is not located at $x=0$.  Using $V_{max}$, we can determine the values for $x_0$ and $x_1$:
\beeq
\begin{array}{ll}
x_0=1+\ln \left[  \frac{V_{max}}{l(l+1)+(1-s^2)}  \right]~,\\ \\
x_1=\sqrt{\dfrac{l(l+1)}{V_{max}}}  ~.
\label{x0x1}
\end{array}
\eneq

The solution to the Regge-Wheeler equation (\ref{RWEnoTim}) for the potential (\ref{VIpotential}) is
\beeq
\phi_l(x) = \left\{ \begin{array}{ll}
	A  (-1)^{-i\omega}\Gamma(1-2i\omega)I_{-2i\omega}\left(2\sqrt{\left[l(l+1)+(1-s^2)\right] e^{x-1}}\right) \\ 
	 ~~+   B (-1)^{i\omega}\Gamma(1+2i\omega) I_{2i\omega}\left(2\sqrt{\left[l(l+1)+(1-s^2)\right] e^{x-1}}\right)& \mbox{$x < x_0$}~\\  \\
	C e^{i\sqrt{\omega^2-V_{max}}~x} + D e^{-i\sqrt{\omega^2-V_{max}}~ x} & \mbox{$x_0 \le x < x_1$}~\\    \\
	E e^{i\omega x}\left( -1+\frac{3}{\omega^2 x^2}+ \frac{3}{i\omega x} \right) + F  e^{-i\omega x}\left( -1+\frac{3}{\omega^2 x^2}- \frac{3}{i\omega x} \right)  & \mbox{$x \geq x_1$~,}
\end{array}
\right.        
\label{solution}
\eneq
where $\omega$ is the complex QNM frequency to be determined.  $I_{\pm \alpha}(z)$ are the modified Bessel functions of the first kind.  $A$, $B$, $C$, $D$, $E$, and $F$ are constants.  The boundary conditions (\ref{B.C.}) require that $B=F=0$.

We determine the QNM frequencies by applying the following conditions at the two points $x_0$ and $x_1$:
\beeq
\begin{array}{ll}
	\phi_l(x< x_0)|_{x=x_0} = \phi_l(x> x_0)|_{x=x_0}~\\
    \phi_l'(x< x_0)|_{x=x_0} = \phi_l'(x> x_0)|_{x=x_0}~\\
    \phi_l(x< x_1)|_{x=x_1} = \phi_l(x> x_1)|_{x=x_1}~\\
    \phi_l'(x< x_1)|_{x=x_1} = \phi_l'(x> x_1)|_{x=x_1}~,
\end{array}       
\label{DiscontinuityEq1}
\eneq
where prime indicates derivative with respect to $x$.

To obtain a better understanding of how each segment of the potential (\ref{VIpotential}) contributes to the QNM frequency spectrum, we separate the potential $V_I(x)$ into the following three potentials:
\beeq
V_{Ia}(x) = \left\{ \begin{array}{ll}
	\left[l(l+1)+(1-s^2)\right]e^{x-1}  & \mbox{$x < x_0$}~\\
     0  & \mbox{$x \geq x_0$~,}
\end{array}
\right.        
\label{V1}
\eneq
\beeq
V_{Ib}(x) = \left\{ \begin{array}{ll}
	0  & \mbox{$x < x_0$}~\\
	V_{max}  & \mbox{$x_0 \le x < x_1$}~\\
	0 & \mbox{$x \geq x_1$~,}
\end{array}
\right.        
\label{V2}
\eneq
\beeq
V_{Ic}(x) = \left\{ \begin{array}{ll}
	0  & \mbox{$x < x_1$}~\\
	l(l+1)/x^2  & \mbox{$x \geq x_1$~.}
\end{array}
\right.        
\label{}
\eneq
We then determine the QNM frequencies for these potentials following a similar procedure to that used for potential $V_I$.

For potential $V_{Ia}(x)$, the solution in region $x\ge x_0$ is of the form $c_1 e^{-i\omega x}+c_2 e^{i\omega x}$, where $c_1$ and $c_2$ are constants. $c_1=0$ due to the boundary condition at infinity.  We then can apply the first two conditions in Eq.\ (\ref{DiscontinuityEq1}) and find the QNM frequencies.

In the case of $V_{Ib}$, after imposing the boundary conditions, we find the solution in region $x < x_0$ to be  $c_1 e^{-i\omega x}$ and in region $x \ge x_0$ to be $c_2 e^{i\omega x}$.  We then apply the four conditions in Eq.\ (\ref{DiscontinuityEq}) to determine the QNMs.

Finally, for $V_{Ic}(x)$, the solution that is consistent with boundary conditions (\ref{B.C.}) in region $x < x_1$ is $c_1 e^{-i\omega x}$.  We apply the last two conditions in Eq.\ (\ref{DiscontinuityEq1}) to obtain the QNM frequencies.

We provide the first eleven QNMs for potentials $V_{Ia}$,  $V_{Ib}$, $V_{Ic}$ and $V_I$ in Table II for scalar perturbations ($s=0$) with $l=2$.  In the case of potential $V_{I}$, the QNM frequencies separate into two branches.  One branch lines up along the imaginary axis and the other along the real axis.  It is not difficult to link these two branches to $V_{Ia}$ and $V_{Ib}$ respectively.

\vspace{0.5cm}
\footnotesize
\begin{tabular}{cccccc}
	\multicolumn{6}{c}{Table II:  First eleven QNMs of the potentials $V_{Ia}$, $V_{Ib}$, $V_{Ic}$ and $V_{I}$ for $s=0$ and $l=2$} \\ 
	\hline
	$n$ & $V_{Ia}$ & $V_{Ib}$ & $V_{Ic}$ & 1st Branch: $V_{I}$  & 2nd Branch: $V_{I}$ \\ 
	\hline 
	0 & $0.3868-0.4698i$ & $1.8478-0.6729i$ & $0.5091-0.3424i$ & $1.0717-0.1491i$ &  \\ 
	1 & $0.0000-1.0601i$ & $3.6145-1.1380i$ & $0.0000-0.5329i$ & $1.3468-0.5883i$ &  \\ 
	2 & $0.0000-1.4943i$ & $5.4468-1.3917i$ &  & $0.0000-1.0868i$ &  \\ 
	3 & $0.0000-2.0002i$ & $7.2863-1.5663i$ &  & $0.5904-1.6035i$ & $2.0153-1.1809i$ \\ 
	4 & $0.0000-2.5000i$ & $9.1268-1.6998i$  &  & $0.0000-2.2079i$ & $2.9289-1.5734i$ \\ 
	5 & $0.0000-3.0000i$ & $10.9672-1.8081i$ &  & $0.0000-2.4582i$ & $3.8681-1.8339i$ \\ 
	6 & $0.0000-3.5000i$ & $12.8071-1.8992i$ &  & $0.0000-3.0017i$ & $4.8072-2.0297i$ \\ 
	7 & $0.0000-4.0000i$ & $14.6467-1.9779i$ &  & $0.0000-3.4999i$ & $5.7436-2.1872i$ \\ 
	8 & $0.0000-4.5000i$ & $16.4859-2.0472i$ &  & $0.0000-4.0000i$ & $6.6775-2.3195i$ \\ 
	9 & $0.0000-5.0000i$ & $18.3247-2.1091i$ &  & $0.0000-4.5000i$ & $7.6091-2.4335i$ \\ 
	10 & $0.0000-5.5000i$ & $20.1633-2.1650i$ &  & $0.0000-5.0000i$ & $8.5389-2.5340i$ \\ 	
	\label{Table2}
\end{tabular} 
\normalsize

For better visualization, we plot the data of Table II in Figure \ref{VIgraph}.  For comparison, we also plot the QNM data for $V_{l=2}^{scalar}$.  As mentioned earlier, the 2nd branch of $V_I$ can be linked visually to the QNMs of $V_{Ib}$.  For damping rates of $|\omega_I|\gtrsim 2.5$, the first branch of $V_I$ closely follows the QNMs of $V_{Ia}$.  The link between the roots of $V_I$ and $V_{Ia}$/$V_{Ib}$/$V_{Ic}$ becomes less obvious for the QNMs with lower values of $|\omega|$.

\begin{figure}[th!]
	\begin{center}
		\includegraphics[height=5cm]{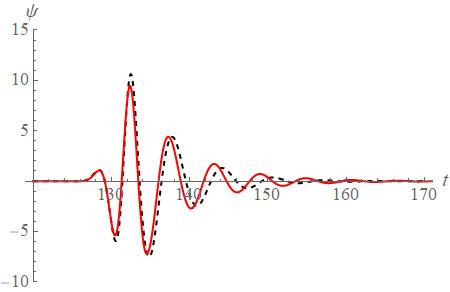}
		\includegraphics[height=5cm]{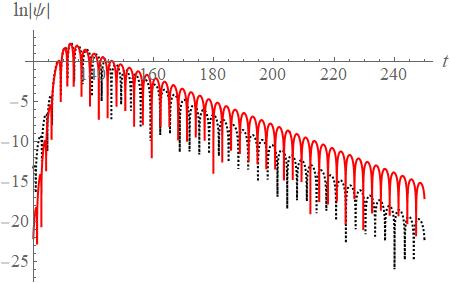}
	\end{center}
    \vspace{-0.5cm}
	\caption{\footnotesize $\psi$ (left) and $\ln |\psi|$ (right), shown in solid red, as a function of time for the potential $V_I$.  In both graphs, for comparison, the ringdown waveform for $V_{l=2}^{scalar}$ is included in dashed black.}
	\label{VI-ringdown}
\end{figure}

The ringdown waveform of the potential $V_I$ is shown in Figure \ref{VI-ringdown}.  In the same figure, for comparison, we show the ringdown waveform for $V_{l=2}^{scalar}$.  The waveform caused by the potential $V_I$ is clearly less damped.  This is consistent with the fact that the fundamental QNM of the potential $V_I$ (Table II) is less damped than the fundamental QNM of $V_{l=2}^{scalar}$ (Table I).

 
\subsection{Potential II}

\begin{figure}[th!]
	\begin{center}
        \includegraphics[height=7.5cm]{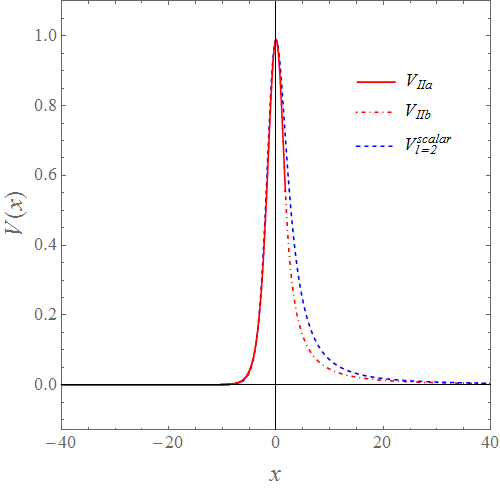}
		\includegraphics[height=8cm]{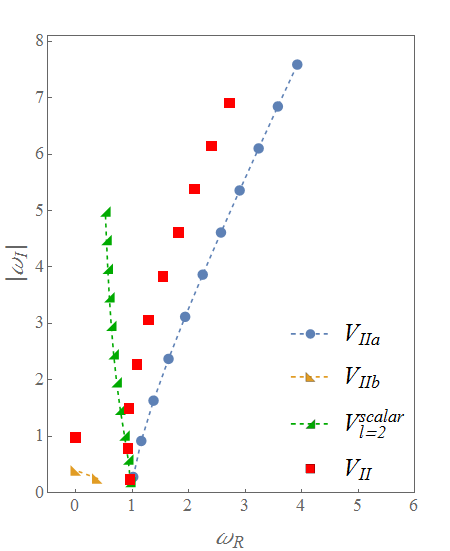}
	\end{center}
    \vspace{-0.5cm}
	\caption{\footnotesize Potential $V_{II}$ is plotted on the left.  The solid red line is $V_{IIa}$ and the dotted-dashed red line is  $V_{IIb}$.  $V_{IIa}$ and $V_{IIb}$ are tangent to each other at $x_0=1.7320$.  $V_{l=2}^{scalar}$ is plotted in dashed blue for comparison.  The QNM data of Table III is presented on the right.  For comparison, the QNM spectrum of $V_{l=2}^{scalar}$ is also included. }
	\label{VIIgraph}
\end{figure}

The second potential we consider involves an inverted P\"{o}schl-Teller potential\cite{Poschl-Teller1, Poschl-Teller2}:
\beeq
V_{II}(x) = \left\{ \begin{array}{ll}
	\dfrac{V_{max}}{\cosh^2(\kappa x)}  & \mbox{$x < x_0$}~\\  \\
	\dfrac{l(l+1)}{(x-a)^2}  & \mbox{$x \geq x_0$~,}
\end{array}
\right.        
\label{VPTx2}
\eneq
where $V_{max}$ is the height of the P\"{o}schl-Teller potential and $\kappa$ and $a$ are free parameters. We show potential (\ref{VPTx2}) in Figure \ref{VIIgraph}.  In the same figure, we also include the potential $V_{l=2}^{scalar}$ for comparison.

The QNM frequencies of the P\"{o}schl-Teller potential can be found in \cite{Starinets, Cardona}:
\beeq
\omega_n=\sqrt{V_{max}-\frac{\kappa^2}{4}}-i \kappa \left(n+\frac{1}{2}\right)~,~n=0,1,2,\dots .
\label{PTw}
\eneq  
We choose $V_{max}$ to be equal to the height of the Regge-Wheeler potential.  We then determine $\kappa$ by requiring that the real part of the QNM frequency of the P\"{o}schl-Teller potential coincide with the real part of the fundamental QNM frequency of the Regge-Wheeler potential. 

The free parameter $a$ can be used to move the function $ l(l+1)/(x-a)^2$ to the right or left.  We adjust $a$ so that the two pieces of the potential in (\ref{VPTx2}) are tangent to each other at $x_0$.

The solution to the Regge-Wheeler equation (\ref{RWEnoTim}) for the potential (\ref{VPTx2}) is
\beeq
\phi_l(x) = \left\{ \begin{array}{ll}
	A  e^{-i\omega x}(1+e^{2\kappa x})^{\beta}~   _2F_1(\beta, \beta-i\omega/\kappa, 1-i\omega/\kappa; -e^{2\kappa x}) \\ 
	~~+  	B  e^{i\omega x}(1+e^{2\kappa x})^{\beta} ~_2F_1(\beta, \beta+i\omega/\kappa, 1+i\omega/\kappa; -e^{2\kappa x})& \mbox{$x < x_0$}~\\  \\

	C e^{i\omega x}\left( -1+\frac{3}{\omega^2 (x-a)^2}+ \frac{3}{i\omega (x-a)} \right) + D  e^{-i\omega x}\left( -1+\frac{3}{\omega^2 (x-a)^2}- \frac{3}{i\omega (x-a)} \right)  & \mbox{$x \geq x_0$~,}
\end{array}
\right.        
\label{solutionPT}
\eneq
where 
\beeq
\beta=\dfrac{1}{2}\left( 1+\sqrt{1-4\dfrac{V_{max}}{\kappa^2}}\right)~.
\label{}
\eneq  
$_2F_1(a,b,c;z)$ is the hypergeometric function.  $A$, $B$, $C$, and $D$ are constants.  Applying the boundary conditions (\ref{B.C.})
indicates that $B=D=0$ in Eq.\ (\ref{solutionPT}).

We determine the QNM frequencies by applying the first two conditions in (\ref{DiscontinuityEq1}) at the point $x_0$.  The solutions are given in Table III.

To obtain a better understanding of how each segment of the potential (\ref{VPTx2}) contributes to the QNM frequency spectrum, we separate the potential $V_{II}$ into the following two potentials:
\beeq
V_{IIa}(x) = \left\{ \begin{array}{ll}
		\dfrac{V_{max}}{\cosh^2(\kappa x)}   & \mbox{$x < x_0$}~\\ \\
	0  & \mbox{$x \geq x_0$~,}
\end{array}
\right.        
\label{VIIa}
\eneq
\beeq
V_{IIb}(x) = \left\{ \begin{array}{ll}
	0  & \mbox{$x < x_0$}~\\ \\
	\dfrac{l(l+1)}{(x-a)^2}  & \mbox{$x \geq x_0$~.}
\end{array}
\right.        
\label{VIIb}
\eneq
We then determine the QNM frequencies for these potentials following the procedure used above for $V_{II}$.

In the case of $V_{IIa}$, after imposing the boundary conditions at $x=\infty$, we find the solution in region $x \ge x_0$ to be  $c_1 e^{i\omega x}$.  We then apply the first two conditions in Eq.\ (\ref{DiscontinuityEq1}) to determine the QNM frequencies.

For $V_{IIb}(x)$, the solution that is consistent with boundary conditions (\ref{B.C.}) in region $x < x_0$ is $c_1 e^{-i\omega x}$.  Once again, we apply the first two conditions in (\ref{DiscontinuityEq1}) to obtain the QNM frequencies.

\vspace{0.5cm}
\footnotesize
\begin{tabular}{ccccc}
	\multicolumn{4}{c}{Table III:  First eleven QNMs of the potentials $V_{IIa}$, $V_{IIb}$ and $V_{II}$} for $l=2$\\ 
	\hline
	$n$ & $V_{IIa}$ & $V_{IIb}$ & $V_{II}$   \\ 
	\hline 
	0 & $1.0161-0.2839i$ & $0.3816-0.2566i$ & $0.9600-0.2400i$ \\ 
	1 & $1.1589-0.9216i$ & $0.0000-0.3994i$ & $0.9290- 0.7858i$   \\ 
	2 & $1.3768-1.6354i$ &  & $0.0000-0.9819i$ \\ 
	3 & $1.6424-2.3746i$ &  & $0.9439-1.4957i$  \\ 
	4 & $1.9365-3.1203i$ &   & $1.0836-2.2746i$ \\ 
	5 & $2.2479-3.8671i$ &  &  $1.2935-3.0571i$  \\ 
	6 & $2.5708-4.6133i$ & & $1.5428-3.8355i$ \\ 
	7 & $2.9014-5.3585i$ &  & $1.8167-4.6087i$ \\ 
	8 & $3.2379-6.10260i$ &  & $2.1074-5.3770i$ \\ 
	9 & $3.5786-6.8457i$ &  & $2.4099-6.1413i$ \\ 
	10 & $3.9228-7.5879i$ &  & $2.7213-6.9021i$ \\ 
\label{Table3}
\end{tabular} 
\normalsize

\begin{figure}[th!]
	\begin{center}
		\includegraphics[height=5cm]{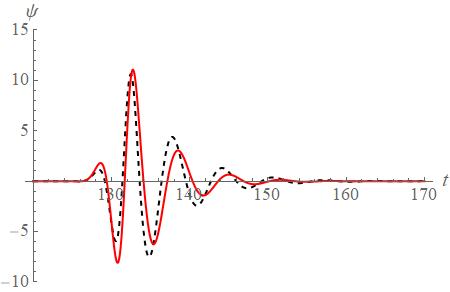}
		\includegraphics[height=5cm]{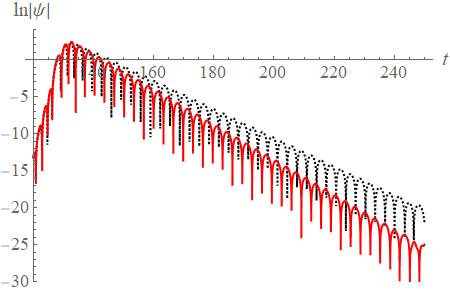}
	\end{center}
	\caption{\footnotesize $\psi$ (left) and $\ln |\psi|$ (right), shown in solid red, as a function of time for the potential $V_{II}$.  In both graphs, for comparison, the ringdown waveform for $V_{l=2}^{scalar}$ is included in dashed black.}
	\label{VII-ringdown}
\end{figure}

We provide the QNMs of the potentials $V_{IIa}$ and $V_{IIb}$ in Table III.  For better visualization, we plot the data of Table III in Figure \ref{VIIgraph}.  For comparison, on the same graph, we also plot the QNM spectrum for $V_{l=2}^{scalar}$.  The QNMs of the potential $V_{II}$ more or less follow the QNMs of the potential $V_{IIa}$ with lower values of $\omega_R$.  We also see a situation similar to the ``algebraically special'' QNM, with a purely imaginary frequency, discussed by Chandrasekhar\cite{Chandra}.  In our case, this special QNM can be clearly linked to $V_{IIb}$.

The ringdown waveform of the potential $V_{II}$ is shown in Figure \ref{VII-ringdown}.  In the same figure, for comparison, we show the ringdown waveform for $V_{l=2}^{scalar}$.  The ringdown waveform caused by the potential $V_{II}$ is more damped.  This is consistent with the fact that the fundamental QNM of the potential $V_{II}$ (Table III) is more damped than the fundamental QNM of $V_{l=2}^{scalar}$ (Table I).


\section{Summary and Conclusion}
\label{Sec:conclusions}

We show that approximating the Regge-Wheeler potential, on a large enough domain, using a piecewise linear function or step functions can lead to an identical ringdown waveform in the limit where the width of each segment of the approximate potential, $\Delta x$, approaches zero.  However, QNMs of the Regge-Wheeler potential, which line up along the imaginary axis, are very different than those of the approximate potentials, which line up along the real axis.   

Using approximate potentials  provides a new mathematical tool that makes calculating the ringdown waveform computationally less intensive than using the Regge-Wheeler potential. Piecewise linear functions approximate the Regge-Wheeler potential much better than step functions, and can be used to generate the waveform faster. However, it is easier to determine the QNMs of step potentials. The difficulty in computing the QNMs of the piecewise linear potential is due to the highly oscillatory behavior of Airy functions.

In \cite{Nollert}, using step functions, Nollert was hoping to find a QNM spectrum that achieves two goals. First, that the spectrum should contain individual modes that dominate the ringdown waveform,  and second, that a whole set of QNMs should completely describe the entire time evolution of the waveform.  
The resulting set of QNMs in \cite{Nollert} appear to form a complete set, which satisfies the second goal, but the first goal was not attained.  See \cite{complete1, complete2, complete3, complete4} for an in depth discussion on the completeness of black hole QNMs.  In addition, Nollert concluded that step functions were not able to generate the very late time behavior of the ringdown waveform.  According to Nollert's analysis, this is because the fundamental QNM of the approximate potential, which dominates the very late time behavior of the waveform, is less damped than the fundamental QNM of the Regge-Wheeler potential.  However, we show that the ringdown waveform generated by the approximate potentials can get arbitrarily close to the ringdown of the Regge-Wheeler potential, even at very late times.
In fact, in the case of a piecewise linear potential, a $\Delta x$ of $0.1$ is small enough to generate a waveform that, for all practical purposes, is indistinguishable from the exact waveform.  We show that the apparent disagreement in the late time waveforms noticed by Nollert is due to the appearance of echoes caused by coarse approximations.

We also experiment with smoother approximations to Regge-Wheeler.  Using these potentials, we are able to show that different QNMs can be linked to different regions of a potential.  Small changes in the shape of the potential can have a noticeable impact on the ringdown waveform and an even more significant impact on the QNM spectrum.  This may be useful for testing any alternative/quantum gravity model that leads to a modification in the shape of the QNM potential.  More speculatively, if spacetime is quantized at some microscopic level, we may expect that the actual QNMs of black holes are very different than the QNMs of the Regge-Wheeler potential and perhaps resemble those of the discrete approximations presented in this paper.

While the ringdown waveforms in the time domain for the Regge-Wheeler and approximate potentials look almost identical in Figure \ref{Vlin-ringdown} for $\Delta x=0.1$ and Figure \ref{Vstep-ringdown} for $\Delta x=0.0001$, a more rigorous analysis of the differences of the two waveforms, using the method of matched-filtering, could be employed to understand the extent of their similarity.  Details of using the matched-filtering method in the study of black hole parameters can be found in \cite{Berti-Cardoso-Will}.
 
\vskip .5cm

\leftline{\bf Acknowledgments}
We thank Hans-Peter Nollert for sharing with us some of his numerical code for computing QNMs and Craig Calcaterra for useful discussions.



\def\jnl#1#2#3#4{{#1}{\bf #2} #3 (#4)}

\def\Zphys{{\em Z.\ Phys.} }
\def\jssc{{\em J.\ Solid State Chem.\ }}
\def\jpsJ{{\em J.\ Phys.\ Soc.\ Japan }}
\def\ptps{{\em Prog.\ Theoret.\ Phys.\ Suppl.\ }}
\def\PTP{{\em Prog.\ Theoret.\ Phys.\  }}
\def\LNC{{\em Lett.\ Nuovo.\ Cim.\  }}

\def\JMP{{\em J. Math.\ Phys.} }
\def\NPB{{\em Nucl.\ Phys.} B}
\def\NP{{\em Nucl.\ Phys.} }
\def\PLB{{\em Phys.\ Lett.} B}
\def\PL{{\em Phys.\ Lett.} }
\def\PRL{\em Phys.\ Rev.\ Lett. }
\def\PRA{{\em Phys.\ Rev.} A}
\def\PRB{{\em Phys.\ Rev.} B}
\def\PRD{{\em Phys.\ Rev.} D}
\def\PR{{\em Phys.\ Rev.} }
\def\PRe{{\em Phys.\ Rep.} }
\def\AP{{\em Ann.\ Phys.\ (N.Y.)} }
\def\RMP{{\em Rev.\ Mod.\ Phys.} }
\def\ZPC{{\em Z.\ Phys.} C}
\def\SCI{\em Science}
\def\CMP{\em Comm.\ Math.\ Phys. }
\def\MPLA{{\em Mod.\ Phys.\ Lett.} A}
\def\IJMPA{{\em Int.\ J.\ Mod.\ Phys.} A}
\def\IJMPB{{\em Int.\ J.\ Mod.\ Phys.} B}
\def\cmp{{\em Com.\ Math.\ Phys.}}
\def\JPA{{\em J.\  Phys.} A}
\def\CQG{\em Class.\ Quant.\ Grav.~}
\def\ATMP{\em Adv.\ Theoret.\ Math.\ Phys.~}
\def\AJP{\em Am.\ J.\ Phys.~}
\def\PRSA{{\em Proc.\ Roy.\ Soc.\ Lond.} A }
\def\ibid{{\em ibid.} }
\vskip 1cm

\leftline{\bf References}

\renewenvironment{thebibliography}[1]
        {\begin{list}{[$\,$\arabic{enumi}$\,$]}  
        {\usecounter{enumi}\setlength{\parsep}{0pt}
         \setlength{\itemsep}{0pt}  \renewcommand{\baselinestretch}{1.2}
         \settowidth
        {\labelwidth}{#1 ~ ~}\sloppy}}{\end{list}}


\end{document}